\documentclass[a4paper,11pt]{article}
\usepackage{pos}
\usepackage{amsmath}
\usepackage{bbold}

\newcommand{\tens}[1]{%
  \mathbin{\mathop{\otimes}\limits_{#1}}%
}

\newcommand{\ltapprox}{\raisebox{-0.3ex}{$\,\stackrel{<}{\scriptstyle\sim}\,$}}

\title{Inclusion of heavy spin effects in the $u d \bar{b} \bar{b}$ $I(J^{P})=0(1^{-})$ four-quark channel in the Born-Oppenheimer approximation}
\ShortTitle{Inclusion of heavy spin effects in the $u d \bar{b} \bar{b}$ $I(J^{P})=0(1^{-})$  four quark channel}

\author*[a]{Jakob Hoffmann}
\author[b]{André Zimermmane-Santos}
\author[a,c]{Marc Wagner}

\affiliation[a]{Johann Wolfgang Goethe-Universit\"at Frankfurt am Main, Institut f\"ur Theoretische Physik, Max-von-Laue-Stra{\ss}e 1, D-60438 Frankfurt am Main, Germany}

\affiliation[b]{Deutsches Elektronen-Synchrotron DESY, Notkestra{\ss}e 85, D-22607 Hamburg, Germany}

\affiliation[c]{Helmholtz Research Academy Hesse for FAIR, Campus Riedberg, Max-von-Laue-Stra{\ss}e 12, \\ D-60438 Frankfurt am Main, Germany}

\emailAdd{jhoffmann@itp.uni-frankfurt.de}
\emailAdd{andre.zimermmane@desy.de}
\emailAdd{mwagner@itp.uni-frankfurt.de}

\abstract{We refine our previous study of a $u d \bar{b} \bar{b}$ tetraquark resonance with quantum numbers $I(J^{P})=0(1^{-})$, which is based on antiheavy-antiheavy lattice QCD potentials, by including heavy quark spin effects via the mass difference of the $B$ and the $B^{*}$ meson.
This leads to a coupled channel Schr\"odinger equation, where the two channels correspond to $BB$ and $B^{*}B^{*}$, respectively.
We search for $\mbox{T}$ matrix poles in the complex energy plane, but do not find any indication for the existence of a tetraquark resonance in this refined coupled channel approach.
We also vary the antiheavy-antiheavy potentials as well as the $b$ quark mass to further understand the dynamics of this four-quark system.
}

\FullConference{%
 The 39th International Symposium on Lattice Field Theory,
 8th-13th August, 2022,
Rheinische Friedrich-Wilhelms-Universität Bonn, Bonn, Germany
}

\begin{document}
\maketitle


\section{Introduction}

Antiheavy-anitheavy-light-light tetraquarks have been studied extensively in recent years and received a lot of interest, in particular since there is experimental evidence that such states exist in the form of the $u d \bar c \bar c$ tetraquark $T_{cc}$ discovered by LHCb \cite{LHCb:2021vvq,LHCb:2021auc}. 

On the theoretical side there is strong evidence that tetraquarks with flavor $u d \bar b \bar b$ and $I(J^P) = 0(1^+)$ as well as with flavor $u s \bar b \bar b$ and $J^P = 1^+$ exist and are strong-interaction-stable, even though they have not been discovered by experiments yet. Within lattice QCD these systems were studied using antiheavy-antiheavy potentals and the Born-Oppenheimer approximation \cite{Bicudo:2012qt,Bicudo:2015vta,Bicudo:2015kna,Bicudo:2016ooe} and also by full simulations, where Non Relativistic QCD is used for the heavy $\bar b$ quarks \cite{Francis:2016hui,Junnarkar:2018twb,Leskovec:2019ioa,Hudspith:2020tdf,Mohanta:2020eed}.

In Ref.\ \cite{Bicudo:2017szl} we predicted a $u d \bar b \bar b$ tetraquark resonance with quantum numbers $I(J^P) = 0(1^-)$ around $17 \, \text{MeV}$ above the $B B$ threshold. There we used a rather simple single channel Born-Oppenheimer setup, where effects due to the heavy quark spins are neglected. Since such effects could be of the order $\mathcal{O}(m_{B^\ast} - m_B) = \mathcal{O}(45 \, \text{MeV})$, it is essential to include these effects in a refined study of this tetraquark resonance, which we found in Ref.\ \cite{Bicudo:2017szl} close to the $B B$ threshold with a rather large width around $112 \, \text{MeV}$. In this work, we carry out such a refined study by using an approach \cite{Bicudo:2016ooe} developed in the context of the strong-interaction-stable $I(J^P) = 0(1^+)$ tetraquark, which is also based on the Born-Oppenheimer approximation, but where two coupled channels, $B B$ and $B^* B^*$, are considered.


\section{Theoretical basics}

All calculations presented in this work are based on the Born-Oppenheimer approximation, which is composed of two major steps.

In the first step the heavy $\bar{b}$ quarks are treated as static and temporal correlation matrices of antiheavy-antiheavy-light-light interpolators are computed using standard methods from lattice QCD (see e.g.\ Ref.\ \cite{Bicudo:2015kna}). From these matrices corresponding antiheavy-antiheavy potentials can be extracted.

For isospin $I = 0$ suitable interpolators generating $u d \bar{Q} \bar{Q}$ four-quark states are given by
\begin{equation}
\label{EQN001} O_{BB}=(C\mathbb{L})_{\alpha\beta} (C\mathbb{S})_{\gamma\delta} \Big(\bar{Q}^{a}_{\gamma}(\vec{r}_{1})u^{a}_{\alpha}(\vec{r}_{1})\Big) \Big(\bar{Q}^{b}_{\delta}(\vec{r}_{2})d^{b}_{\beta}(\vec{r}_{2})\Big) - (u \leftrightarrow d) ,
\end{equation}
where $\alpha, \beta, \gamma, \delta$ are spin indices, $a, b$ are color indices, $\vec{r}_{1}$ and $\vec{r}_{2}$ denote the positions of the static antiquarks $\bar{Q}$, $\mathbb{L}$ and $\mathbb{S}$ are $4 \times 4$ spin matrices and $C = \gamma_0 \gamma_2$ is the charge conjugation matrix. Since static spins do not appear in the Hamiltonian and are conserved quantities, it is important to couple the two static spins separately from the two light spins using $\mathbb{S}$ and $\mathbb{L}$, respectively. In the static limit there are only $4$ linearly independent heavy spin matrices $\mathbb{S} \in \{ (\mathbb{1}+\gamma_0) \gamma_5 , (\mathbb{1}+\gamma_0) \gamma_j \}$, for which $O_{BB}$ does not vanish. 
For the light spin coupling there are $16$ possibilities, $\mathbb{L} \in \{ (\mathbb{1}+\gamma_0) , (\mathbb{1}+\gamma_0) \gamma_5 , (\mathbb{1}+\gamma_0) \gamma_j , (\mathbb{1}+\gamma_0) \gamma_5 \gamma_j \}$. In total there are, thus, $4 \times 16 = 64$ interpolators $O_{BB}$, which correspond to the $64$ possible combinations of pairs of $B$, $B^{*}$, $B_{0}$ and $B_{1}^{*}$ mesons.

\begin{figure}
\begin{center}
\includegraphics[width=9.0cm]{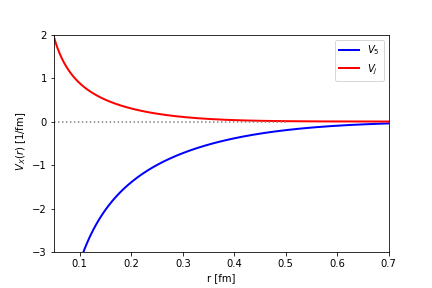}
\end{center}
\caption{\label{FIG001}Parametrizations of lattice QCD results for the $I=0$ antiheavy-antiheavy potentials $V_5$ and $V_j$ as functions of the separation $r$.}
\end{figure}

For the Schr\"odinger equation discussed below we need potentials associated with pairs of $B$ and $B^{*}$ mesons, since channels containing a $B_{0}$ or a $B_{1}^{*}$ meson decouple. There are two such potentials, a strongly attractive potential $V_5$ corresponding to $\mathbb{L} = (\mathbb{1}+\gamma_0) \gamma_5$ and a weakly repulsive potential $V_j$ corresponding to $\mathbb{L} \in \{ (\mathbb{1}+\gamma_0) \gamma_j \}$. Lattice QCD results extrapolated to physically light $u/d$ quark masses can be parameterized consistently by
\begin{equation}
\label{EQN010} V_x(r) = -\frac{\alpha_x}{r} e^{-(r/d_x)^2} \quad , \quad x = 5,j
\end{equation}
with $\alpha_5=0.34_{-0.03}^{+0.03}$, $d_5=0.45_{-0.10}^{+0.12} \, \text{fm}$ \cite{Bicudo:2015kna} and $\alpha_j=-0.10 \pm 0.07$, $d_j=(0.28 \pm 0.017) \, \text{fm}$ \cite{Bicudo:2016ooe}. The parameterizations are shown in Figure~\ref{FIG001}.

In the second step of the Born-Oppenheimer approximation the potentials $V_{5}$ and $V_{j}$ are used in a $16 \times 16$ coupled channel Schr\"odinger equation, where each channel corresponds to a particular pair of $B$ and $B^*$ mesons. Effects due to the heavy quark spins are introduced via the mass splitting $m_{B^{*}}-m_{B} \approx 45 \, \text{MeV}$ taken from experiments \cite{ParticleDataGroup:2022pth}.

In detail, the Schr\"odinger equation for the relative coordinate of the two $\bar b$ quarks is given by
\begin{equation}
\label{EQN004} H \psi(\vec{r}) = E \psi(\vec{r}) ,
\end{equation}
where, at large $\bar b \bar b$ separations $|\vec{r}|$, the 16 components of the wave function represent meson-meson pairs according to
\begin{eqnarray}
\nonumber & & \hspace{-0.7cm} \psi \equiv (B B     , B B_x^*     , B B_y^*     , B B_z^*     \ \ , \ \
B_x^* B , B_x^* B_x^* , B_x^* B_y^* , B_x^* B_z^* \ \ , \ \
B_y^* B , B_y^* B_x^* , B_y^* B_y^* , B_y^* B_z^* \ \ , \\
\label{EQN003} & & \hspace{0.675cm} B_z^* B , B_z^* B_x^* , B_z^* B_y^* , B_z^* B_z^*)^T
\end{eqnarray}
(the indices $x,y,z$ refer to the orientation of the spins of the $B^*$ mesons). $H$ is the Hamilton operator,
\begin{equation}
H = M\tens{}\mathbb{1}_{4\times4} + \mathbb{1}_{4\times4}\tens{}{M} + \frac{\vec{p}^2}{2 \mu} + H_{\text{int}}
\end{equation}
with the meson masses $M=\text{diag}(m_{B},m_{B^{*}},m_{B^{*}},m_{B^{*}})$, the relative momentum of the the two $\bar b$ quarks and the reduced $b$ quark mass $\mu = m_b / 2$. $H_{\text{int}}$ is a $16 \times 16$ non-diagonal matrix containing the antiheavy-antiheavy potentials $V_5$ and $V_j$ computed in the first step. It is given by
\begin{equation}
H_{\text{int}} = T V_{\text{diag}} T^{-1} ,
\end{equation}
where $V_{\text{diag}}$ is a diagonal $16 \times 16$ matrix with entries $V_5$ or $V_j$ and $T$ is a $16 \times 16$ transformation matrix, whose entries relate antiheavy-antiheavy potentials to meson-meson pairs. $T$ can be determined using Fierz identities, i.e.\ by expressing the fermion bilinears $(C\mathbb{L})_{\alpha\beta} u^{a}_{\alpha} d^{b}_{\beta}$ and $(C\mathbb{S})_{\gamma\delta} \bar{Q}^{a}_{\gamma} \bar{Q}^{b}_{\delta}$ in Eq.\ (\ref{EQN001}) in terms of the 16 components of the wave function defined in Eq.\ (\ref{EQN003}) (for details we refer to Ref.\ \cite{Bicudo:2016ooe}). We note again that there is no coupling between channels with $B_{0}$ and/or $B_{1}^{*}$ mesons and the $16$ channels listed in Eq.\ (\ref{EQN003}).

One can show that both orbital angular momentum $L$ as well as total spin $S$ are conserved quantities. This allows to decompose the $16 \times 16$ equation (\ref{EQN004}) into independent $1 \times 1$ and $2 \times 2$ equations with either symmetric or antisymmetric wave functions with respect to meson exchange.
In this work we are exclusively interested to study a possibly existing resonance with $I(J^P) = 0(1^-)$ predicted in Ref.\ \cite{Bicudo:2017szl}, which has $L = 1$ and $S = 0$. Thus, we focus on the $2 \times 2$ block representing $S = 0$,
\begin{eqnarray}
\label{EQN005} \left( \begin{pmatrix} 2m_{B} && 0 \\ 0 && 2m_{B^{*}} \end{pmatrix} + \frac{\vec{p}^2}{2 \mu} + V_{2 \times 2}(r) \right)\vec{\psi}_{2 \times 2}(\vec{r}) = E \vec{\psi}_{2 \times 2}(\vec{r}) .
\end{eqnarray}
The potential matrix $V_{2 \times 2}$ contains linear combinations of the potentials $V_{5}$ and $V_{j}$,
\begin{equation}
\label{EQN012} V_{2 \times 2}(r) =\frac{1}{4} \begin{pmatrix} V_{5}(r)+3V_{j}(r) && \sqrt{3} (V_{5}(r)-V_{j}(r)) \\ \sqrt{3} (V_{5}(r)-V_{j}(r)) && 3V_{5}(r)+V_{j}(r) \end{pmatrix} ,
\end{equation}
and entries of the 2-component wave function $\vec{\psi}_{2 \times 2}$ can be interpreted according to
\begin{equation}
\label{EQN006} \vec{\psi}_{2 \times 2} \equiv ( BB \ \ , \ \ \vec{B}^{*} \vec{B}^{*} / \sqrt{3} ) .
\end{equation}


\section{Scattering formalism}

In this section we specialize the Schr\"odinger equation (\ref{EQN005}) to $L = 1$. Moreover, we use standard techniques from scattering theory to implement boundary conditions, which define the $\mbox{T}$ matrix (see e.g.\ Refs.\ \cite{Bicudo:2017szl,Bicudo:2020qhp} for more detailed discussions).

We write each of the $2$ components of the wave function (\ref{EQN006}) as a superposition of an incoming partial wave proportional to $A_\alpha j_1(k_\alpha r)$, $\alpha  \in \{ B B , B^* B^* \}$, which is a solution of the free Schr\"odinger equation, and an emergent spherical wave proportional to $\chi_\alpha(k_\alpha r) / r$, i.e.\
\begin{equation}
\vec{\psi}_{2 \times 2}(\vec{r}) = \left(\begin{array}{c}
A_{B B}           j_1(k r)   + \chi_{B B}(r) / r \\
A_{B^\ast B^\ast} j_1(k^* r) + \chi_{B^\ast B^\ast}(r) / r
\end{array}\right) Y_{1,m}(\vartheta,\varphi)
\end{equation}
with $k = k_{B B} = \sqrt{2 \mu (E - 2m_B)}$ and $k^* = k_{B^* B^*} = \sqrt{2 \mu (E-2m_{B^*})}$. This leads to an ordinary differential equation for the radial coordinate $r$,
\begin{eqnarray}
\nonumber & & \hspace{-0.7cm} \left(\begin{pmatrix} 2 m_B && 0 \\ 0 && 2 m_{B^{*}} \end{pmatrix} - \frac{1}{2 \mu} \left(\frac{d^2}{dr^2} - \frac{2}{r^2}\right) + V_{2 \times 2}(r) - E\right) \begin{pmatrix} \chi_{B B}(r) \\ \chi_{B^* B^*}(r) \end{pmatrix} = \\
\label{EQN013} & & = -\frac{r}{4} \begin{pmatrix}
 (V_5(r) + 3 V_j(r)) A_{B B} j_1(k r) + \sqrt{3} (V_5(r) - V_j(r)) A_{B^* B^*} j_1(k^* r) \\
 \sqrt{3} (V_5(r) - V_j(r)) A_{B B} j_1(k r) + (3 V_5(r) + V_j(r)) A_{B^* B^*} j_1(k^* r)
 \end{pmatrix} .
\end{eqnarray}
$A_{B B}$ and $A_{B^* B^*}$ reflect the composition of the incoming partial wave in terms of a $B B$ and a $B^* B^*$ meson pair, respectively. Moreover, the emergent wave is proportional to the corresponding scattering amplitude and $h_1^{(1)}(k_\alpha r)$ for large $r$. This allows to define the $2 \times 2$ $\mbox{T}$ matrix,
\begin{equation}
\label{EQN007} \mbox{T} = \begin{pmatrix}
t_{B B;B B} && t_{B B;B^* B^*} \\
t_{B^* B^*;B B} && t_{B^* B^ *;B^* B^*}
\end{pmatrix} ,
\end{equation}
via
\begin{eqnarray}
\label{EQN008} & & \hspace{-0.7cm} \chi_\alpha(r) = i r t_{B B;\alpha} h_1^{(1)}(k_\alpha r)_{{}^{\phantom{* *}}} \quad \text{for } r \rightarrow \infty \text{ and } (A_{B B} , A_{B^* B^*}) = (1,0) \\
\label{EQN009} & & \hspace{-0.7cm} \chi_\alpha(r) = i r t_{B^* B^*;\alpha} h_1^{(1)}(k_\alpha r) \quad \text{for } r \rightarrow \infty \text{ and } (A_{B B} , A_{B^* B^*}) = (0,1) .
\end{eqnarray}
The first index of $t_{\alpha,\beta}$ refers to the incoming wave, the second index to the emergent wave. Poles of the $\mbox{T}$ matrix indicate bound states (for $\text{Re}(E) < 2m_B$) and resonances (for $\text{Re}(E) > 2m_{B}$).


\section{\label{SEC004}Numerical results}

We used a standard fourth order Runge-Kutta method to solve the Schr\"odinger equation (\ref{EQN013}) for given complex energy $E$ and determined the corresponding $\mbox{T}$ matrix (\ref{EQN007}) by comparing the wave function to the boundary conditions (\ref{EQN008}) and (\ref{EQN009}) at large $r$. Then we applied a Newton-Raphson root finding algorithm to $1/\det(\mbox{T}(E))$ to determine the poles of the $\mbox{T}$ matrix in the complex energy plane.

When using the potentials (\ref{EQN010}), meson masses $m_B = 5280 \, \text{MeV}$, $m_{B^*} = 5325 \, \text{MeV}$ from the PDG \cite{ParticleDataGroup:2022pth} and $m_b = 4977 \, \text{MeV}$ from a quark model \cite{Godfrey:1985xj}, we did not find a pole. In particular, there is no sign of the $I(J^P) = 0(1^-)$ $u d \bar b \bar b$ tetraquark resonance predicted in Ref.\ \cite{Bicudo:2017szl} within a more basic single channel setup, where heavy spin effects and the mass splitting of the $B$ and the $B^*$ meson are neglected. The reason, why this resonance does not exist in the more realistic coupled channel setup introduced in this work, seems to be the competition between the attractive potential $V_5$ and the repulsive potential $V_j$. In the lighter $B B$ channel the potential is $(V_5 + 3 V_j) / 4$ (see the upper left element of the potential matrix (\ref{EQN012})), i.e.\ the attractive contribution is suppressed by the factor $1/4$, disfavoring the existence of a resonance. In the $B^* B^*$ channel the potential is $(3 V_5 + V_j) / 4$ (see the lower right element of the potential matrix (\ref{EQN012})), i.e.\ the attractive component still dominates, but the potential is shifted upwards by $2 (m_{B^*} - m_B) = 90 \, \text{MeV}$, which is also not favorable concerning the formation of a resonance.


\subsection{Varying the potential $V_j$}

To understand this in more detail we replaced the repulsive potential $V_j$ from Eq.\ (\ref{EQN010}) by $V_j = \epsilon V_5$ and varied $\epsilon$ in the range $-0.50 \leq \epsilon \leq +1.00$. $\epsilon = -0.50$ is quite similar to $V_j$ from Eq.\ (\ref{EQN010}) (see Figure~\ref{FIG001}; this is also supported by the leading order of perturbation theory, where $V_j = -V_5/2$), while $\epsilon = +1.00$ leads to a decoupling of the two channels, i.e.\ to a setup containing the single channel equation used in Ref.\ \cite{Godfrey:1985xj}. Thus, continuously changing $\epsilon$ from $+1.00$ to $-0.50$ allows to study the existence and properties of the resonance predicted in Ref.\ \cite{Bicudo:2017szl}, while smoothly transitioning from the single channel setup of Ref.\ \cite{Bicudo:2017szl} to the coupled channel setup used in this work.

\begin{figure}
\begin{center}
\includegraphics[width=9.0cm]{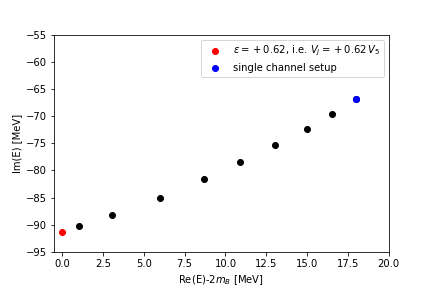}
\includegraphics[width=6.0cm]{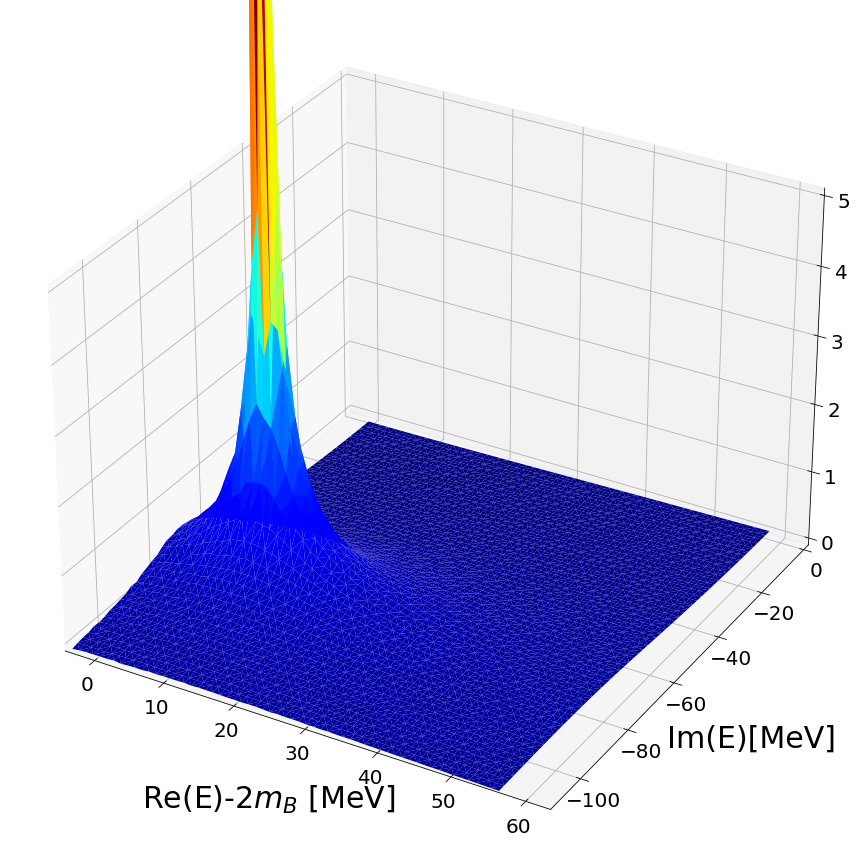}
\end{center}
\caption{\label{FIG002}
\textbf{(left)}~Trajectory of the $\mbox{T}$ matrix pole in the complex energy plane, when decreasing $\epsilon$ from $+1.00$ to $+0.62$, where the pole disappears.
\textbf{(right)}~$|\text{det}(\mbox{T})|$ as function of the complex energy for $\epsilon = +0.83$.
}
\end{figure}

In the left plot of Figure~\ref{FIG002} we show the trajectory of the $\mbox{T}$ matrix pole in the complex energy plane, when decreasing $\epsilon$ from $+1.00$ (blue point, single channel setup from Ref.\ \cite{Bicudo:2017szl}\footnote{The energy of the $\epsilon = 1.00$ data point in Figure~\ref{FIG002} is slightly different from the energy quoted in Ref.\ \cite{Bicudo:2017szl}, because in that reference a different value for $m_b$ was used.}) to $+0.62$ (red point, still an attractive potential $V_j = +0.62 \times V_5$), where the pole disappears. Since $\epsilon = +0.62$ is rather far away from $\epsilon = -0.50$ (which is close to the lattice QCD result for $V_j$), our results clearly suggest that the $I(J^P) = 0(1^-)$ resonance does not exist in nature.

The right plot of Figure~\ref{FIG002} is an exemplary plot of $|\text{det}(\mbox{T})|$ as function of the complex energy for $\epsilon = +0.83$, i.e.\ a strongly attractive $V_j = +0.83 \times V_5$. There is a clear pole $\text{Re}(E) - 2 m_B = 12.4 \, \text{MeV}$ above the $B B$ threshold. The width of the corresponding resonance is $\Gamma = -2 \, \text{Im}(E) = 153 \, \text{MeV}$.

\begin{figure}
\begin{center}
\includegraphics[width=9.0cm]{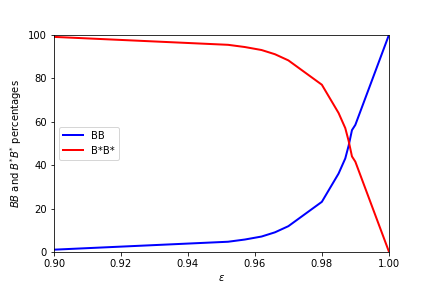}
\end{center}
\caption{\label{FIG003}Meson composition of the resonance as function of $\epsilon$.}
\end{figure}

To further illustrate the interplay between the potentials $V_5$ and $V_j$ on the one hand and the mass difference $m_{B^*} - m_B$ on the other hand, we computed the meson composition of the resonance as function of $\epsilon$ using a technique proposed in Ref.\ \cite{Bicudo:2020qhp}. For given $\epsilon$ we determine the energy of the $\mbox{T}$ matrix pole and solve the Schr\"odinger equation (\ref{EQN013}) for real energy $\text{Re}(E_\text{pole})$. The squares of the emergent spherical waves are proportional to the probabilities to find the system in a $B B$ or in a $B^* B^*$ state, respectively. Thus the meson percentages can be defined as
\begin{equation}
\% \alpha = \frac{\alpha}{B B + B^* B^*} \quad , \quad \alpha = B B , B^* B^*
\end{equation}
with
\begin{equation}
B B = \int_0^{R_\text{max}} dr \, \Big|\chi_{B B}(r)\Big|^2 \quad , \quad B^* B^* = \int_0^\infty dr \, \Big|\chi_{B^* B^*}(r)\Big|^2 .
\end{equation}
$\% B B$ and $\% B^* B^*$, computed for $R_\text{max} = 2.0 \, \text{fm}$ (the meson percentages are only weakly dependent on $R_\text{max}$), are shown in Figure~\ref{FIG003}. One can see that, at $\epsilon = +1.00$, the resonance is a pure $B B$ state. This is expected, because the two channels decouple and the lighter $B B$ channel is identical to the single channel setup from Ref.\ \cite{Bicudo:2017szl} and, thus, fully contains the predicted resonance. When decreasing $\epsilon$, the $B B$ and the $B^* B^*$ channels mix and the system selects dynamically the energetically most favorable meson composition. In other words, the lighter but less attractive $B B$ channel competes with the heavier but more attractive $B^* B^*$ channel. We found that there is a rather quick transition from $\% B B = 100 \%$ (and $\% B^* B^* = 0 \%$) at $\epsilon = 1.00$ to $\% B B < 10 \%$ (and $\% B^* B^* > 90 \%$) for $\epsilon < 0.96$. The obvious interpretation is that an attractive potential is significantly more important for the formation of the resonance than lighter meson constituents.

\subsection{Heavier than physical $b$ quark masses}

We also explored heavier than physical $b$ quark masses by defining
\begin{eqnarray}
 & & \hspace{-0.7cm} m_b = \kappa m_{b,\text{phys}} \\
\label{EQN016} & & \hspace{-0.7cm} m_{B^*} - m_B = \frac{m_{B^*,\text{phys}} - m_{B,\text{phys}}}{\kappa} = \frac{45 \, \text{MeV}}{\kappa}
\end{eqnarray}
and performing computations with $\kappa > 1$.
$m_{b,\text{phys}} = 4977 \, \text{MeV}$ is the quark model value for the $b$ quark mass already used in the previous subsection and Eq.\ (\ref{EQN016}) represents the leading order Heavy Quark Effective Theory prediction for the mass splitting of the $B$ and the $B^*$ meson \cite{Neubert:1996wg}. 

\begin{figure}
\begin{center}
\includegraphics[width=9.0cm]{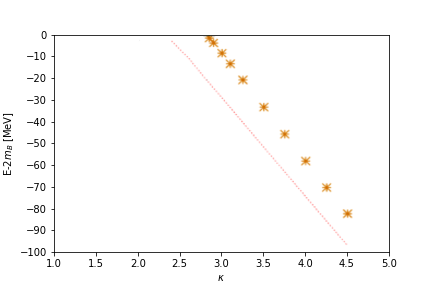}
\end{center}
\caption{\label{FIG004}Binding energy $E - 2 m_B$ as function of $\kappa$ (yellow dots: coupled channel setup; dotted line: single channel setup from Ref.\ \cite{Bicudo:2017szl}).}
\end{figure}

Since the potentials are independent of $\kappa$, there must be a bound state for sufficiently large $m_b$. This expectation is confirmed by numerical results, where a bound state appears at $\kappa \approx 2.8$. This can be seen in Figure~\ref{FIG004}, where we show the binding energy $E - 2 m_B$ as function of $\kappa$. For comparison, we also show corresponding results obtained in the single channel setup of Ref.\ \cite{Bicudo:2017szl}. The single channel results are similar, but shifted to smaller values of $\kappa$ and a bound state already appears at $\kappa \approx 2.4$.

There is, however, a significant qualitative difference between results obtained in the single channel setup and the coupled channel setup of this work for $1 < \kappa \ltapprox 2.4$ and $1 < \kappa \ltapprox 2.8$, respectively. In the single channel setup there is a resonance, but in the coupled channel setup a resonance does not seem to exist, not even when $\kappa$ is approaching $2.8$, i.e.\ for $m_b$ close to that value, where a bound state appears. We plan to investigate this in more detail in the near future.

\section{Conclusions}

Our results presented in section~\ref{SEC004} suggest that a $u d \bar b \bar b$ tetraquark resonance with $I(J^P) = 0(1^-)$ does not exist. We note, however, that our approach, even though based on lattice QCD potentials, is not fully rigorous and resorts to certain approximations, e.g.\ the Born-Oppenheimer approximation. When comparing the mass of the strong-interaction-stable $u d \bar b \bar b$ tetraquark with $I(J^P) = 0(1^+)$ obtained in the same approach \cite{Bicudo:2016ooe} to results from recent full lattice QCD computations \cite{Francis:2016hui,Junnarkar:2018twb,Leskovec:2019ioa,Hudspith:2020tdf,Mohanta:2020eed}, there is a discrepancy in the binding energy, around $59 \, \text{MeV}$ versus $100 \ldots 150 \, \text{MeV}$, which is not fully understood yet. Since full lattice QCD leads to stronger binding, it cannot be excluded, that the resonance exists in full QCD. Thus, to complement the results presented in this work, it would be interesting to also explore the $I(J^P) = 0(1^-)$ sector using full lattice QCD with methods similar to those from Refs.\ \cite{Leskovec:2019ioa,Meinel:2022lzo}.


\section*{Acknowledgements}

We acknowledge useful discussions with Pedro Bicudo, Lasse M\"uller, Martin Pflaumer and Jonas Scheunert.
J.H.\ and M.W.\ acknowledge support by the Deutsche Forschungsgemeinschaft (DFG, German Research Foundation) -- project number 457742095. A.Z.-S.\ acknowledges support by a Goethe Goes Global Scholarship. M.W.\ acknowledges support by the Heisenberg Programme of the Deutsche Forschungsgemeinschaft (DFG, German Research Foundation) -- project number 399217702.
Calculations were conducted on the GOETHE-HLR and on the FUCHS-CSC high-performance computers of the Frankfurt University. We would like to thank HPC-Hessen, funded by the State Ministry of Higher Education, Research and the Arts, for programming advice.



\end{document}